\documentclass[pra,twocolumn]{revtex4}
\usepackage{graphicx,color}
\usepackage{epsfig}
\usepackage{graphicx}
\usepackage{mathrsfs}
\usepackage{color}
\usepackage{bm}
\usepackage{amsmath,amssymb,amsthm,amscd}
\usepackage{mdwlist}

\newcommand{\Tr}{\operatorname{Tr}}
\begin{document}

\title{Optimal convex approximations of quantum states}
\author{Massimiliano F. Sacchi}
\affiliation{Istituto di Fotonica e Nanotecnologie - CNR, Piazza Leonardo
  da Vinci 32, I-20133, Milano, Italy}
\affiliation{QUIT group, Dipartimento di Fisica, 
Universit\`a di Pavia, via A. Bassi 6, I-27100 Pavia, Italy}
\date{\today}

\begin{abstract} 
We consider the problem of optimally approximating an
unavailable quantum state $\rho $ by the convex mixing of states drawn
from a set of available states $\{ \nu _i\}$.  The problem is recast
to look for the least distinguishable state from $\rho $ among the
convex set $\sum _i p_i \nu _i$, and the corresponding optimal weights
$\{ p_i \}$ provide the optimal convex mixing.  We present the
complete solution for the optimal convex approximation of a qubit
mixed state when the set of available states comprises the three bases
of the Pauli matrices.  
\end{abstract}

\maketitle 
\section{Introduction}
Convex structures are ubiquitous in the realm of quantum mechanics.
Density matrices, probability operator-valued
measures, and completely positive maps---which represent quantum states, 
quantum measurements, and quantum channels, respectively---are convex sets. 
The weights in a convex sum typically 
represent classical probabilities which have an immediate operational 
interpretation: they are the weights of the extremal points of the set and may correspond 
to classical processing. 

In Ref. \cite{ss} the problem of optimally approximating an
unavailable quantum channel by the convex mixing of channels which are
supposed to be available was addressed.  This operational problem has
been recast to the problem of looking for the least distinguishable
channel from the target among the convex set of channels constructed
by the given set.
 
Here in this paper we address the analogous problem for quantum states, namely the 
problem of optimally generating a desired quantum state $\rho $, when only a given set of quantum
states $\{\nu  _i \}$ 
is disposable.  In this case, 
we will look for the best convex combination among
the states of the given set that mostly resembles the desired $\rho $, 
i.e. that is the least distinguishable from $\rho $ itself.  
As for the case of convex approximation of channels, this
approach has clearly a prompt experimental application when 
the effectively available states in a lab are limited
for intrinsic restrictions, unavailable technology, or even economical reasons.  
A further relevance of this approach is due to the fact that a convex sum of
states offers the possibility of performing  different
experiments followed by post-processing of experimental data 
when the quantities of interest are linear with respect to the input states. 

Since the natural measure of distinguishability between quantum states is based on the 
trace norm \cite{hel}, 
we note that our general problem of convex approximation includes the well-studied 
(and still open) problem of quantifying the coherence of quantum states for the 
specific case where the available set $\{\nu  _i \}$ corresponds to a complete orthogonal 
basis, via the trace-distance measure of coherence \cite{coh,coh2,coh3}. 
Also, for generic mutipartite state $\rho $ 
and available set of states given by all product pure states, 
our problem is equivalent to evaluate the trace norm of entanglement measure 
\cite{enta,enta2,enta3,enta4}.

\section{Convex approximation of quantum states}
It is well known that the probability $p_{\mbox{\scriptsize discr}}$  of optimally discriminating 
between two quantum states $\rho_0 $ and $\rho  _1$ given with equal {\em a priori} probability 
is given by \cite{hel}
\begin{eqnarray}
p_{\mbox{\scriptsize discr}}
(\rho _0, \rho _1) =\frac 12 +\frac 14 ||\rho  _0 -\rho  _1|| _1
\;,\label{uno}
\end{eqnarray}
where $\Vert  A \Vert _1 $ denotes the trace norm of $A$, namely \cite{bhatia} 
\begin{eqnarray}
\Vert A \Vert _1= \hbox{Tr}\sqrt{A^\dag A}= \sum _i s_i (A)\;, 
\end{eqnarray}
$\{s_i (A)\}$ representing the singular values of $A$. In the case of
Eq. (\ref{uno}), 
the singular values just correspond to the absolute value of the
eigenvalues, since the operator inside the norm is Hermitian. Let us also recall that  
the optimal measurement for the discrimination is performed by 
the projectors on the support of the positive and negative 
part of the Hermitian operator $\rho_0 -\rho _1$. 

The problem of the optimal convex approximation of a quantum state is implicitly posed by 
the following definition. 

{\bf Definition:} The optimal convex approximation of a quantum state
$\rho $ with respect to (w.r.t.) a given set of quantum states $\{ \nu  _i \}$ is
given by $\sum _ i p_i ^{opt} \nu  _i$, where $\{ p_i^{opt}\}$
denotes the vector of probabilities 
\begin{eqnarray} \{p_i ^{opt}\}=
\arg \min _ {\{ p_i\}} \Vert \rho - \textstyle\sum _i p_i \nu  _i
\Vert _1 \;.  \end{eqnarray} 
The effectiveness of the optimal convex
approximation is then quantified by the $\{ \nu  _i \}$-{\em
distance} 
\begin{eqnarray} D_{\{\nu  _i \}}(\rho )\equiv \min _ {\{
p_i\}} \Vert \rho - \textstyle\sum _i p_i \nu  _i \Vert _1 \;,
\end{eqnarray} 
which provides the worst probability of discriminating
the desired state $\rho $ from any of the available states $\sum _ i
p_i \nu  _i$.  Clearly, our definition of optimal convex
approximation can be suitably changed by referring to any other figure
of merit that quantifies the distance between quantum states (e.g., a decreasing function of 
the fidelity).

We notice that the formulation of the trace norm as a semidefinite program \cite{boyd} 
allows its efficient calculation. 
Moreover, the convexity of the norm itself allows one  
the search of the minimum by means of standard software of convex optimization \cite{cvx,cvx2}. 

From the convexity of the trace norm, it follows the upper  bound 
\begin{eqnarray}
D_{\{\nu   _i \}}(\rho )\leq \min _ {i} ||\rho  - \nu   _i ||_1  = \min _i 
D_{ \nu  _i }(\rho  )
\;.\label{upp}
\end{eqnarray}
Notice also that for all  
unitary operators  $U$, from the unitarily invariance of the trace norm,  
one has the symmetry
\begin{eqnarray}
D_{\{ \nu  _i \}}(\rho )=D_{ \{ U \nu  _i U^\dag \} }
(U \rho U^\dag )
\;.  
\end{eqnarray}
Clearly, if the set itself $\{ \nu  _i\}$ is invariant, then  
\begin{eqnarray}
D_{\{\nu  _i \}}(U \rho U^\dag )=
 D_{\{ \nu   _i \}}(\rho )
\;,\label{inva}
\end{eqnarray}
and the probabilities of the optimal convex approximation for  
$U \rho U^\dag $ are just a permutation of those for $\rho  $. 
This is the case, for example, when the set of the available quantum states is covariant w.r.t. 
a (projective) unitary representation of a group.

\section{Pauli distance of qubit states}
In the following we provide the complete analytical solution for the 
optimal convex approximation of an arbitrary mixed qubit state, 
when the available set of states is given by the eigenvectors of the three Pauli matrices. 

Let us first consider the simpler case where 
the set of available states is an orthogonal basis. Without loss of generality let us
identify such basis as the eigenstates ${\sf B}_1= \{|0 \rangle ,|1
\rangle \} $ of $\sigma _z $-Pauli matrix, and parametrize the target qubit state $\rho $ as
\begin{eqnarray} 
\rho =\left( 
\begin{array}{cc} 1-a 
& k 
\sqrt{a(1-a)}e^{-i\phi }\\ 
k \sqrt{a(1-a)}e^{i\phi } 
& a  \\ \end{array} \right ) \;,\label{rho} 
\end{eqnarray} with
$a \in [0,1]$, $\phi \in [0,2\pi]$, and $k\in [0, 1]$.
A straightforward calculation provides the optimal convex approximation
of $\rho $ as the diagonal matrix 
\begin{eqnarray} \rho _d=\left(
\begin{array}{cc} 1-a  & 0 \\ 0 & a
\\ 
\end{array} \right ) 
\;.\label{uabd} 
\end{eqnarray}
Clearly, the optimal weights are given by $p ^{opt }_0=1-a$ and
$p^{opt} _1 =a$, and the approximation is
quantified by the ${\sf B}_1$-distance 
\begin{eqnarray} D_{\{|0 \rangle ,|1
\rangle \} }(\rho )= 2 k \sqrt{a(1-a)}
\;. 
\end{eqnarray} 
This result also corresponds to the trace-distance measure of coherence for the state $\rho $ 
referred to the $\sigma _z$ eigenstates \cite{cohq}. 
Better approximations can be obviously obtained when a larger set of states is available. 

Let us consider now the set containing the eigenstates of all
Pauli matrices, namely 
\begin{eqnarray}
{\sf B}_3  = \!\!\!\!\!&&
\left \{    |0 \rangle , |1
\rangle , |2 \rangle \equiv \frac{1}{\sqrt 2} (|0 \rangle +|1 \rangle ), |3
\rangle \equiv \frac {1}{\sqrt 2} 
(|0 \rangle - |1 \rangle ),  
\right. 
\nonumber \\
&& \left.  |4 \rangle \equiv \frac {1}{\sqrt 2} 
(|0 \rangle +i |1 \rangle ), |5 \rangle \equiv \frac {1}{\sqrt 2} (|0
\rangle - i |1 \rangle ) 
\right \}
\;.
\end{eqnarray}
Since the ${\sf B}_3$-distance (or, equivalently, {\em Pauli distance}) 
is invariant for the state transformations $
\rho (a,k,\phi)\rightarrow \rho (1-a,k,\phi)$ and 
$\rho (a,k,n  \pi /2 \pm \phi)\rightarrow \rho (a,k,\phi)$ (with integer $n$), 
we can restrict the study to the case $a \in [0,1/2]$ and 
$\phi \in [0,  \pi /2]$. 
One can immediately find a large set of density matrices which indeed correspond 
to a convex mixing of the six states of ${\sf B}_3$.  
In fact, we can rewrite $\rho $ in Eq. (\ref{rho}) as follows 
\begin{eqnarray}
\rho &=& (1-2a)|0 \rangle \langle 0| \nonumber \\& + & 
 2k\sqrt{a(1-a)}(\cos \phi |2 \rangle \langle 2| +
\sin \phi |4 \rangle \langle 4|)\nonumber \\& +& 
[a-k\sqrt{a(1-a)}(\cos \phi +\sin \phi) ] I
\;,\label{ide}
\end{eqnarray}
where $I$ denotes the two-dimensional identity matrix. It follows that 
there is a threshold value for the coherence parameter $k$ under which $D_{{\sf B}_3}(\rho )=0$, 
namely for
\begin{eqnarray}
k\leq k_{th}\equiv \frac {a}{\sqrt{a(1-a) }(\cos\phi +\sin \phi)}\;. \label{kth}
\end{eqnarray}
The pertaining weights that provide such an exact convex decomposition can be chosen as follows \cite{notuni}
\begin{eqnarray}
&&
p_0=1-a-k\sqrt{a(1-a)}(\cos \phi +\sin \phi)\nonumber \\& & 
p_1=a-k\sqrt{a(1-a)}(\cos \phi +\sin \phi)\nonumber \\& & 
p_2=2k \sqrt{a(1-a)}\cos\phi \nonumber \\& & 
p_4=2k \sqrt{a(1-a)}\sin\phi \nonumber \\& & 
p_3=p_5=0 
\;.
\end{eqnarray}
In terms of the expectation values $\langle \sigma _\alpha \rangle =\Tr [\rho \sigma _\alpha ]$, with $\alpha=x,y,z$ and $\rho $ 
as in Eq. (\ref{rho}),  notice the identities
\begin{eqnarray}
&&\langle \sigma _x \rangle =2k \sqrt {a(1-a)}\cos \phi  \;,\nonumber \\ 
&&\langle \sigma _y \rangle =2k \sqrt {a(1-a)}\sin \phi \;, \nonumber \\ 
&&\langle \sigma _z \rangle =1-2a 
\;.
\end{eqnarray}
Thus, the condition $k\leq k_{th}$ in Eq. (\ref{kth}) can be rewritten more transparently as 
\begin{eqnarray}
\langle \sigma _x \rangle + \langle \sigma _y \rangle + \langle \sigma _z \rangle \leq 1
\;.
\end{eqnarray}
With the help of symbolic computation, by imposing  a vanishing value to the gradient of 
$\Vert \rho - \textstyle \sum _{i=0} ^{5} p_i |i \rangle \langle i| \Vert _1$ with respect to the probabilities $\{ p_i \}$, 
one can obtain the complete 
analytical solution for the optimal convex approximation of $\rho $ when $k> k_{th}$, and hence  $D_{{\sf B}_3}(\rho )>0$. 
Explicitly, one obtains the following three cases \cite{nota}: 
\par \noindent  $i)$ 
for $k_{th} < k \leq  \frac {a}{\sqrt{a(1-a)}}$, 
or $k> \frac {a}{\sqrt{a(1-a)}}$ and $\phi \in [\phi _{th},\pi/2 -\phi _{th}]$, with 
\begin{eqnarray}
\!\phi _{th}\!=2 \arctan \!
\left [ \frac{\sqrt{5k^2 a(1-a) -a^2}-2k \sqrt{a(1-a)}}{a+k\sqrt{a(1-a)}} \right ]
\end{eqnarray}
the optimal convex approximation has Pauli distance 
\begin{eqnarray}
D_{{\sf B}_3}(\rho )= \frac {2}{\sqrt 3} \sqrt{a(1-a)(1+\sin 2\phi)}(k-k_{th}) 
\;,\label{duno}
\end{eqnarray}
with pertaining optimal weights
\begin{eqnarray}
&&p_0=1-\frac {4}{3}a  -\frac 23 k \sqrt{a(1-a)} (\cos \phi +\sin \phi) \nonumber 
\\& & 
p_2=\frac 23 [a+k\sqrt {a(1-a)}(2\cos \phi -\sin \phi)]\nonumber \\& & 
p_4=\frac 23 [a+k\sqrt {a(1-a)}(2\sin \phi -\cos \phi)]\nonumber \\& & 
p_1=p_3=p_5=0
\;;
\end{eqnarray}
\par \noindent  $ii)$ 
for $k> \frac {a}{\sqrt{a(1-a)}}$ and $\phi \in [0,\phi _{th}]$, the optimal convex approximation 
has Pauli distance 
\begin{eqnarray}
D_{{\sf B}_3}(\rho )&= &\{2a [a-2k\sqrt{a(1-a)}\cos\phi \nonumber \\& + & 
k^2(1-a)(2- \cos^2 \phi)]\}^{1/2}
\;,\label{ddue}
\end{eqnarray}
with optimal weights
\begin{eqnarray}
&&p_0=1-a-k\sqrt{a(1-a)\cos\phi }\nonumber \\& & 
p_2=a+k\sqrt{a(1-a)\cos\phi }\nonumber \\& & 
p_1=p_3=p_4=p_5=0
\;;
\end{eqnarray}
\par \noindent  $iii)$ 
for $k> \frac {a}{\sqrt{a(1-a)}}$ and $\phi \in [\pi /2 -\phi _{th},\pi /2]$, 
the optimal convex approximation 
has Pauli distance 
\begin{eqnarray}
D_{{\sf B}_3}(\rho ) &= &\{2a [a-2k\sqrt{a(1-a)}\sin\phi \nonumber \\& +& 
k^2(1-a)(2- \sin^2 \phi)]\}^{1/2}
\;,\label{dtre}
\end{eqnarray}
with optimal weights
\begin{eqnarray}
&&p_0=1-a-k\sqrt{a(1-a)\sin\phi }\nonumber \\& & 
p_4=a+k\sqrt{a(1-a)\sin\phi }\nonumber \\& & 
p_1=p_2=p_3=p_5=0
\;.
\end{eqnarray}
Notice that the exact convex decomposition (when $k\leq k_{th}$) involves four states, whereas 
the optimal convex approximation corresponds to a mixture of three states in case i), 
and just two states in cases ii) and iii). 

In Figs. 1 and 2, respectively, 
we plot the results for the optimal convex approximation of $\rho $ 
versus parameters $a$ and $\phi $ with fixed value of the parameter $k=\frac 23$,  
and versus parameters $a$ and $k$ with fixed value of the phase parameter $\phi =\frac \pi 3$.

\begin{figure}[thb]
  \includegraphics[width=\columnwidth]{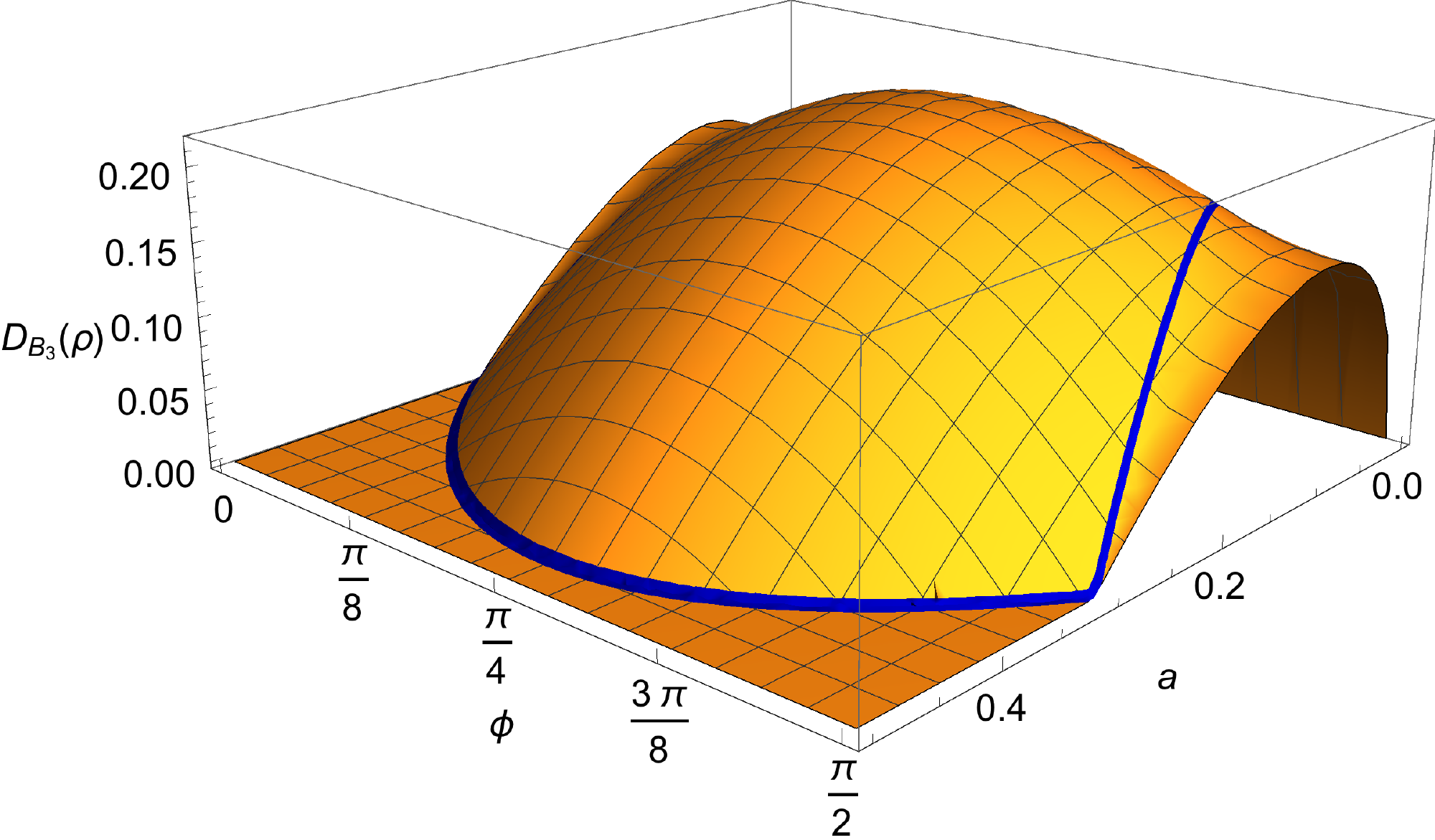}
  \caption{Optimal convex approximation of a qubit mixed state $\rho $ w.r.t. the set $B_3$ of 
the eigenstates of the three Pauli matrices.  The Pauli distance $D_{B_3}(\rho )$ 
is here plotted versus the target state parameters $a$ and $\phi $, 
for fixed value of the parameter $k= \frac 23$ 
[see the parameterization of $\rho$ in Eq. (\ref{rho})]. The plotted surface is a piecewise 
function obtained from Eqs. (\ref{duno}), (\ref{ddue}), and 
(\ref{dtre}) in their region of definition. 
According to Eq. (\ref{kth}),  the minimal trace distance vanishes in the region where  
$\frac {a}{\sqrt{a(1-a) }(\cos\phi +\sin \phi)}\geq k=\frac 23 $.} 
  \label{fig:dvsak}
\end{figure}

\begin{figure}[thb]
\includegraphics[width=\columnwidth]{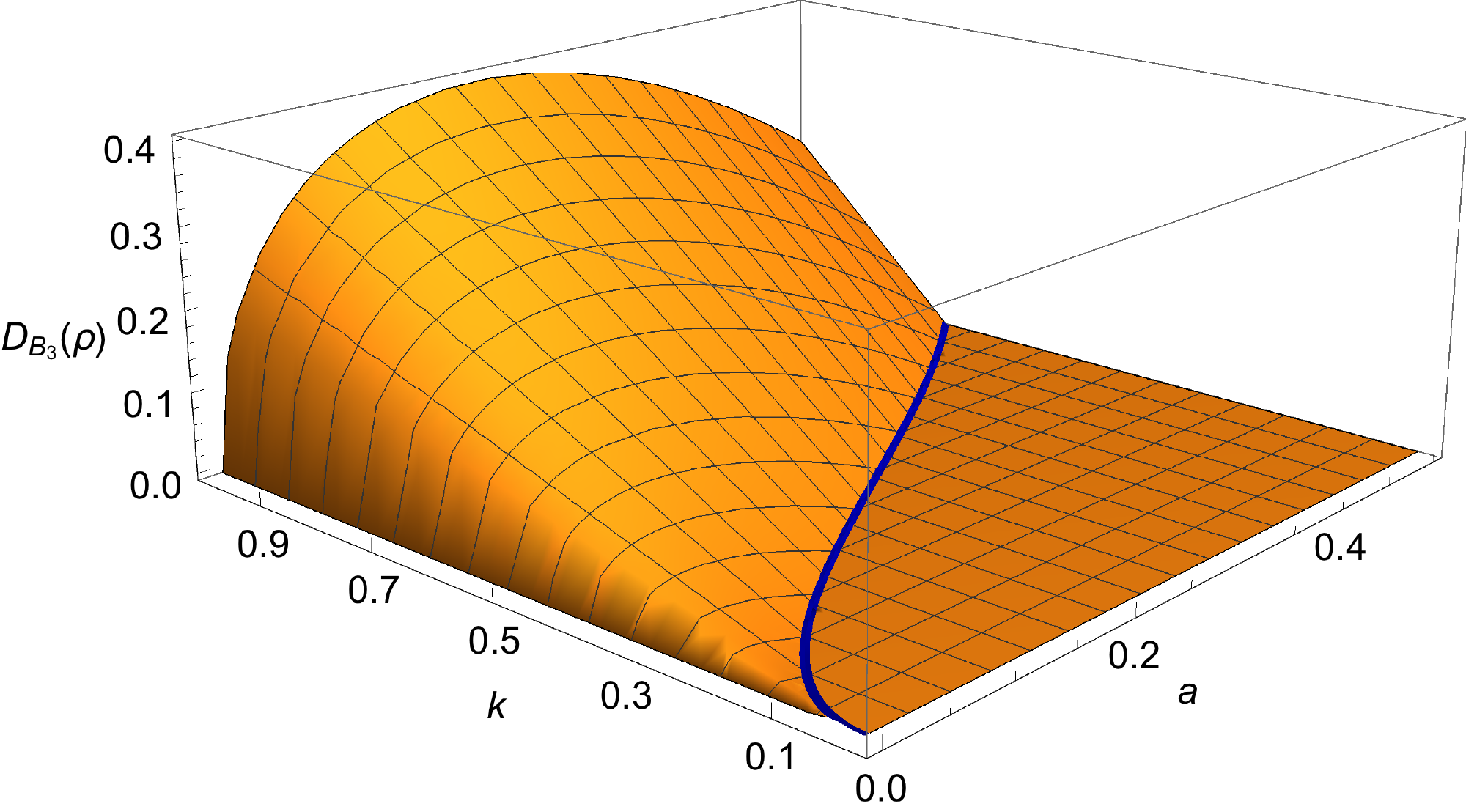}
  \caption{Optimal convex approximation of a qubit mixed state $\rho $ w.r.t. the set $B_3$ of 
the eigenstates of the three Pauli matrices.  The Pauli distance $D_{B_3}(\rho )$ 
is here plotted versus the target state parameters $a$ and $k$, 
for fixed value of the phase $\phi = \frac \pi 3$. 
 According to Eq. (\ref{kth}), 
the minimal trace distance vanishes for $k\leq \frac{2a}{(\sqrt{3}+1)\sqrt{a(1-a)}}$.}
\label{fig:dvsaph}
\end{figure}

\section{Conclusions}
Let us conclude our paper with the following observations. Imagine that we want to 
approximate $N$ copies of the state, namely $\rho ^{\otimes N}$,  
and we have at disposal a set of single-copy states 
$\{\nu  _i \}$. 
The optimal convex approximation in this case provides the distance 
$D_{\{ \otimes _{j=1}^N  \nu  _{i_j} \}}(\rho ^{\otimes N})$. 
Since the convex hull of ${\{ \otimes _{j=1}^N  \nu  _{i_j} \}}$ contains all 
the $N$-fold tensor products $\otimes _{j=1}^N (\sum _i p_{i_j} \nu  _i )$,   
one has 
\begin{eqnarray}
D_{\{\otimes _{j=1}^N  \nu   _{i_j} \}}(\rho  ^{\otimes N}) 
&\leq &\min _{\{p_{i_j} \}} 
\Vert \rho ^{\otimes N} - 
\otimes _{j=1}^N (\textstyle 
\sum _i p_{i_j} \nu  _i ) \Vert _1
\nonumber \\& \leq &
\Vert \rho ^{\otimes N} - 
(\textstyle \sum _i p_i^{opt}  \nu  _i )^{\otimes N} \Vert _1
\;\label{ineq}
\end{eqnarray}
where $\{p_i^{opt}\}$ denotes the vector of probabilities pertaining to the optimal 
convex approximation of a single copy of the state $\rho $. 

We notice that one can find  strict inequalities in both lines of Eq. (\ref{ineq}). 
The first inequality arises because the 
presence of correlations in the 
convex approximation can be beneficial even if the target state is indeed the product 
of independent states (as it occurs, for example, in the optimal cloning of quantum states 
\cite{clon}, where the copies are correlated). In fact, it is also known that correlations 
limit the extractable information \cite{corr,corr2,corr3,corr4}, 
and here indeed we want to minimize the probability of discriminability.  
The second inequality stems from the fact that 
the distance for a convex optimization has no additive/multiplicative 
property with respect to the tensor product. Hence,  even 
looking for a tensor-product state for the optimal convex approximation, 
the corresponding optimal weights will be in general different from those pertaining to  
the optimal convex approximation of a single copy. 
This also implies that we do not have an exact expression for the scaling with $N$ 
of the distance between a quantum state and its convex approximation.
In a systematic study of the scaling of the optimal convex approximation 
with the number of copies, the results related to the quantum 
Chernoff bound  \cite{chern,chern2} might be very useful.  
A specific example where Eq. (\ref{ineq}) is satisfied with two strict inequalities 
is the following. 
Consider the pure qubit state $|\psi \rangle =\frac {\sqrt 3}{2}
|0 \rangle +\frac 12 |1 \rangle $. 
Its optimal convex approximation with respect to the 
basis $\{ |0 \rangle ,|1 \rangle \} 
$ is achieved by $p^{opt}_0=3/4$ and $p^{opt}_1=1/4$, with 
corresponding distance $D_{\{ |0 \rangle ,|1 \rangle \} }(|\psi \rangle \langle \psi |)=\frac 
{\sqrt 3}{2}$. The two-copy trace distance can be evaluated as 
\begin{eqnarray}
\left \Vert |\psi \rangle \langle \psi | ^{\otimes 2} 
- \left ( 
\frac 34 |0 \rangle \langle 0| +\frac 14 
|1 \rangle \langle 1|  \right ) ^{\otimes 2}\right \Vert _1 \simeq 1.299\;. 
\end{eqnarray}

On the other hand, the optimal convex approximation with respect to factorized diagonal states 
is given by 
\begin{align}
& D_{{\sf B}_1 \otimes {\sf B}_1} (|\psi \rangle \langle \psi | ^{\otimes 2} )  =\nonumber \\ 
\min _{p,q\in [0,1]} &
\Vert |\psi \rangle \langle \psi | ^{\otimes 2} - [p |0 \rangle \langle 0| +(1-p) 
|1 \rangle \langle 1|]
\otimes \nonumber \\&  
[q |0 \rangle \langle 0| +(1-q) 
|1 \rangle \langle 1|]
\Vert _1 \simeq 1.272\;, 
\end{align}
and is achieved for $p^{opt}=q^{opt}\simeq 0.859$. 
Finally, by allowing correlations between the copies of the convex approximation, one obtains 
\begin{align}
&D_{\{|00 \rangle ,|01 \rangle ,|10 \rangle ,|11 \rangle \}}(|\psi \rangle \langle \psi | 
^{\otimes 2} )= 
\nonumber \\ \min _ {\{p_{ij} \}} &
\Vert |\psi \rangle \langle \psi | ^{\otimes 2} - 
(p_{00} |00 \rangle \langle 00| +
p_{01} |01 \rangle \langle 01| + 
\nonumber \\& 
 p_{10}
|10 \rangle \langle 10| + p_{11} |11 \rangle \langle 11| )
\Vert _1 \simeq 1.265
\;,
\end{align}
where the optimal weights are given by $p_{00}^{opt}\simeq 0.712$, $p_{01}^{opt}
=p_{10}^{opt}\simeq 0.144$, and $p_{11}^{opt}=0$. Notice that the improvement in the convex 
approximation of $|\psi \rangle \langle \psi | ^{\otimes 2}$ is exclusively due to classical 
correlations, since obviously no entanglement is present in the approximating state.\\

\vspace{0.6cm}

\section{Addendum}
An erroneous parameterization of the phase Eq. (17) has been
propagated to the results in Eqs. (18-23), which are
invalid in some cases. Correct analytical solutions can be found in
Ref. \cite{comm}, which can be viewed as a complete supplement to
Sect. III. As a consequence, also Fig. 2 is inaccurate in some areas. 
We are grateful with the authors of Ref. \cite{comm} 
to have corrected the corresponding results. 

The main observations for the Pauli case still holds, namely that 
i) the exact convex decomposition generally involves four states and
is achieved under the condition of Eq. (13);   
ii) when such a condition is violated, the optimal convex
approximation essentially breaches in two cases, where optimality is
achieved by a mixture of just two or three states. 

All important observations
contained in the conclusions, namely the discussion about the case of
many copies of quantum states, the non-additivity of the distance, and
the role of correlations, maintain their general validity.\\


\begin{thebibliography}{99}
\bibitem{notuni}The solution is not unique, since clearly one can equivalently ascribe 
the weight of the 
identity matrix in Eq. (\ref{ide}) to any of the three Pauli matrix
bases.
\bibitem{nota}The solutions in Eqs. (19-23) are invalid in some
  cases. See the Addendum in Sec. V.
\end{thebibliography}
\end{document}